%% file: main.tex
\documentclass[submission,copyright,creativecommons]{eptcs}
\providecommand{\event}{TARK 2021} 

\usepackage{amssymb}

\usepackage{verbatim}
\usepackage{amsmath}
\usepackage{tikz}
\usepackage{mathrsfs}
\usepackage[margin=10pt,font=small,labelfont=bf]{caption}

\usepackage{graphicx}
\usepackage{dsfont}
\usepackage{enumerate}

\usepackage[title,toc,titletoc,header]{appendix}

\newtheorem{definition}{Definition}

\newtheorem{remark}{Remark}

\newtheorem{example}{Example}

\renewcommand{\phi}{\varphi}

\input{PaperPicture.tex}

\title{A Deontic Stit Logic Based on Beliefs and Expected Utility}

\author{Aldo Iv\'an Ram\'irez Abarca
\institute{Utrecht University\\ Utrecht, The Netherlands}
\email{boiangaleano@hotmail.com}
\and
Jan Broersen
\institute{Utrecht University\\ Utrecht, The Netherlands}
\email{J.M.Broersen@uu.nl}
}

\def\titlerunning{Deontic Stit, Beliefs, EU}
\def\authorrunning{Ram\'irez Abarca \& Broersen}

\begin{document}
\maketitle


\begin{abstract}
The formalization of action and obligation using logic languages is a topic of increasing relevance in the field of ethics for AI. Having an expressive syntactic and semantic framework to reason about agents’ decisions in moral situations allows for unequivocal representations of components of behavior that are relevant when assigning blame (or praise) of outcomes to said agents. Two very important components of behavior in this respect are belief and belief-based action. In this work we present a logic of doxastic oughts by extending epistemic deontic stit theory with beliefs. On one hand, the semantics for formulas involving belief operators is based on probability measures. On the other, the semantics for doxastic oughts relies on a notion of optimality, and the underlying choice rule is maximization of expected utility. We introduce an axiom system for the resulting logic, and we refer to its soundness, completeness, and decidability results. These results are significant in the line of research that intends to use proof systems of epistemic, doxastic, and deontic logics to help in the testing of ethical behavior of AI through theorem-proving and model-checking.
\end{abstract}

\section{Introduction}

It has been argued that an appropriate theory of agency and obligation should take into account what agents know---and what they know how to do---both before and at the moment of acting (\cite{Xu2015}, \cite{horty2017action}, \cite{horty2019epistemic}, \cite{JANANDI}). Following considerations from epistemic game theory (EGT)---which has clear conceptual and technical connections to stit theory (see \cite[Chapter 1]{duijf2018let} and \cite{tamminga2013deontic})---we put forward that agents' beliefs also play an important role in the relation between agency and obligation. In recent years we have seen a growing interest in adding \emph{knowledge} modalities to stit theory, but there are relatively few extensions of this logic by means of \emph{belief} operators (notably those in \cite{wansing2006doxastic} and \cite{broersen2013probabilistic}). The novelty of the present approach lies in its intention to develop a link between beliefs and ought-to-do. This is a natural step to take in the line of both Horty's formalization of \emph{act utilitarian ought-to-do} (\cite{Horty2001}) and its extension with epistemic notions (\cite{horty2019epistemic}, \cite{JANANDI}). 

According to Horty (\cite{Horty2001}), act utilitarian ought-to-do stems from a measure of \emph{optimality} of actions. The consequences of optimal actions are taken to be the conditions that agents ought to have brought about in the world. Horty's idea of \emph{optimality} is undeniably inspired by solution concepts from game theory, particularly by dominance of strategies. Following this idea, the recent works \cite{horty2019epistemic} and \cite{abarca2019logic} introduce \emph{epistemic} (resp. \emph{subjective}) ought-to-do's to account for the relation between knowledge and obligation. To be more precise, in both these works the optimal actions once again underlie what agents epistemically (resp. subjectively) ought to do, but the measure of optimality now takes into consideration agents' epistemic states. The resulting formalization can deal with complex scenarios for which the initial---non-epistemic---ought-to-do fell short of very intuitive standards in the context of responsibility attribution. Since stit theory can---at least in principle---incorporate most game theoretic ideas into multi-agent action-settings, we find it worthy to extend the theory of ought-to-do with a notion of belief. Our long-term goal is to achieve a nuanced formalization of obligation and responsibility, where agents' hierarchies of belief would serve as explanations of the actions that these agents perform interactively. 

Consider the following example, inspired by the famous film \emph{The Verdict}, of 1982. Suppose that a patient is admitted to the hospital in urgent need of surgery.  The nurses draw up a chart with important background information for the surgeons, but unfortunately the figure regarding how long it has been since the patient last ate has a mistake. Anesthetics for this surgery should only be supplied if the patient has had an empty stomach for at least eight hours, and they are deadly otherwise. Because of the mistake in the chart, the anesthesiologist never comes to know that the patient had had a full meal just one hour before admittance. Therefore, she gives the anesthetics and the patient dies. It is clear that the doxastic state of the anesthesiologist plays a key role in determining whether she is morally responsible for the patient's death. On the causal level, the anesthesiologist is responsible for it. However, it seems natural that she should not be held culpable, because she acted upon the false---but justified---belief that the patient had an empty stomach before admittance. Moreover, the anesthesiologist is justified in considering the action of supplying the anesthetics as something that she ought to have done, given the circumstances.  

There are several options of conceptual backgrounds for incorporating beliefs into deontic stit logic (see \cite{wansing2006doxastic}, (\cite{baltag2006conditional}, and \cite{harsanyi1967games} for some alternatives). In this work we adopt a \emph{quantitative} version of agentive belief, and we use probability measures (on the domain of stit structures) to represent doxastic states. An agent's subjective belief in a proposition is taken to depend on the probability that the agent assigns to the indices at which the propositions holds. The reason for choosing a probabilistic semantics of belief is that it allows us to base a notion of \emph{doxastic ought} on a key concept in decision theory: \emph{maximization of expected utility}. Thus, we propose that at a given index of evaluation an agent had the doxastic obligation to see to it that $\phi$ if $\phi$ is a consequence of all the actions that maximized expected (deontic) utility at such index, where this utility is identified with the deontic value of the index just as in \cite{Horty2001}. The basic aspects of the logic that we develop here, then, can be summarized in this way: we extend the basic stit language with (a) epistemic and doxastic operators, (b) objective and subjective ought-to-do operators (whose logic was developed in \cite{abarca2019logic}), and (c) a doxastic ought-to-do operator, meant to build up formulas to characterize the effects of those actions that agents' belief-systems render as optimal---i.e., the effects of rational \emph{best responses} (see \cite{bjorndahl2017reasoning}).

The paper is structured as follows. In Section \ref{shit} we introduce the syntax and semantics of the epistemic deontic logic that we use as basis for our theory of doxastic oughts. In Section \ref{believeyoume} we present the probabilistic conceptualization of belief that we intend to incorporate into the logic of Section \ref{shit}, and we elaborate on the reasons for choosing such a notion of belief. In Section \ref{giventofly} we introduce the syntax and semantics for formulas involving the doxastic ought-to-do operator, and we discuss an example to illustrate its reach within a stit-theoretic analysis of responsiblity and excusability. In Section \ref{axiomatique} we develop an axiomatic system for the resulting logic and address its soundness, completeness, and decidability results, after which we conclude.  

\section{Epistemic Deontic Stit}
\label{shit}

As discussed in \cite{horty2019epistemic}, \cite{JANANDI}, and \cite{abarca2019logic}, an adequate theory of ought-to-do should account for agents' epistemic states before and at the moments of action. This is all the more relevant in contexts of responsibility attribution, and more specifically for excusability (see, for instance, \cite{lorini2014logical}). In principle, if an agent does not know how to fulfill an obligation, it should be excused for not having done so. 
The mentioned \cite{horty2019epistemic}, \cite{JANANDI}, and \cite{abarca2019logic} all extend Horty's stit theory of act utilitarian ought-to-do with knowledge operators and explore the relation between uncertainty and obligation. 
In turn, here we will be extending the logics of \cite{JANANDI} and \cite{abarca2019logic} with modalities for belief.\footnote{The models that we use are simpler and at the same time more general than the ones of \cite{horty2019epistemic}. The same \cite{JANANDI} and \cite{abarca2019logic} discuss some advantages of their models over the ones developed in \cite{horty2019epistemic}.} 

\begin{definition}[Language]
\label{syntax ep stit}
Given a finite set $Ags$ of agent names, a countable set of propositions $P$ such that $p \in P$ and $\alpha \in Ags$, the grammar for the formal language $\mathcal L_{\mathtt{KO}}$ is given by:
\[ \begin{array}{lcl}
\phi :=  p \mid \neg \phi \mid \phi \wedge \psi \mid \square \phi \mid [\alpha] \phi \mid K_\alpha \phi \mid \odot [\alpha] \phi \mid \odot_{\mathcal{S}}[\alpha] \phi
\end{array}. \]
\end{definition}

$\square\varphi$ is meant to express the `historical necessity' of $\varphi$
($\Diamond \varphi$ abbreviates $\neg \square \neg \varphi$). $[\alpha] \varphi$ stands for `agent $\alpha$ has seen to it that $\varphi$.' 
$K_\alpha$ is the epistemic operator for $\alpha$, so that $K_\alpha \phi$ stands for `agent $\alpha$ knows that $\varphi$ holds.'  $\odot [\alpha]\phi$ is meant to express that $\alpha$ objectively ought to have seen to it that $\phi$. Finally, $\odot_{\mathcal{S}}[\alpha] \phi$ is meant to express that $\alpha$ subjectively ought to have seen to it that $\phi$.

As for the semantics, the structures on which we evaluate formulas of the language $\mathcal L_{\mathtt{KO}}$  are based on what we call \emph{epistemic act-utilitarian branching-time frames}. 

\begin{definition}[Epistemic act-utilitarian branching-time frames]
\label{frames}

A \textbf{finite} epistemic act-utilitarian

\noindent branching-time frame (\emph{eaubt}-frame for short) is a tuple $\langle M,\sqsubset,\mathbf{\mathbf{Choice}}, \{\sim_\alpha\}_{\alpha\in Ags}, \mathbf{Value}  \rangle$ such
that:

\begin{itemize}

\item $M$ is a non-empty \emph{finite} set of \textnormal{moments} and $\sqsubset$ is a strict partial ordering on $M$ satisfying `no backward branching.' Each maximal $\sqsubset$-chain is called a $\textnormal{history}$, which represents a way in which time might evolve. $H$ denotes the set of all histories, and for each $m\in M$, $H_m:=\{h \in H ;m\in h\}$. Tuples $\langle m,h \rangle$ are called \emph{indices}  iff $m \in M$, $h \in H$, and $m\in h$.  $\mathbf{Choice}$ is a function that maps each agent $\alpha$ and moment $m$ to a partition $\mathbf{Choice}^m_\alpha$ of $H_m$, where the cells of such a partition represent $\alpha$'s available actions at $m$. For $m\in M$ and $h\in H_m$, we denote the equivalence class of $h$ in $\mathbf{Choice}^m_\alpha$ by $\mathbf{Choice}^m_\alpha(h)$. $\mathbf{Choice}$ satisfies two constraints: $(\mathtt{NC})$ \emph{No choice between undivided histories}: For all $h, h'\in H_m$, if $m'\in h\cap h'$ for some $m' \sqsupset m$, then $h\in L$ iff $h'\in L$ for every $L\in \mathbf{Choice}^m_\alpha$. $(\mathtt{IA})$ \emph{Independence of agency}: A function $s$ on $Ags$ is called a \emph{selection function} at $m$ if it assigns to each $\alpha$ a member of $\mathbf{Choice}^m_\alpha$. If we denote by $\mathbf{Select}^m$ the set of all selection functions at $m$, then we have that for every $m\in M$ and $s\in\mathbf{Select}^m$, $\bigcap_{\alpha \in Ags} s(\alpha)\neq \emptyset$ (see \cite{belnap01facing} for a discussion of the property).

\item For $\alpha\in Ags$, $\sim_\alpha$ is the epistemic indistinguishability equivalence relation for agent $\alpha$, which satisfies the following constraints: $(\mathtt{OAC})$ \emph{Own action condition}: For every index $\langle m_*,h_*\rangle$, if $\langle m_*, h_*\rangle\sim_\alpha \langle m, h\rangle$ for some $\langle m,h\rangle$, then $\langle m_*,h_*'\rangle\sim_\alpha \langle m,h\rangle$ for every $h_*'\in \mathbf{Choice}^{m_*}_\alpha (h_*)$. We refer to this constraint as the `own action condition' because it implies that agents do not know more than what they perform. $(\mathtt{Unif-H})$ \emph{Uniformity of historical possibility}: For every index $\langle m_*,h_*\rangle$, if $\langle m_*, h_*\rangle \sim_\alpha \langle m, h\rangle $ for some $\langle m,h\rangle $, then for every $h_*'\in H_{m_*}$ there exists $h'\in H_m$ such that $\langle m_*, h_*'\rangle \sim_\alpha \langle m, h'\rangle$. Combined with $(\mathtt{OAC})$, this constraint is meant to capture a notion of uniformity of strategies, where epistemically indistinguishable indices should have the same available actions for the agent to choose upon.

For each index $\langle m,h\rangle$ and $\alpha\in Ags$, we define $\alpha$'s \emph{information set} at $\langle m,h\rangle$ as $\pi_\alpha[\langle m, h \rangle]:=\{\langle m', h' \rangle; \langle m, h \rangle\sim_\alpha\langle m', h' \rangle \}$. 

\item $\mathbf{Value}$ is a deontic function that assigns to each history $h\in H$ a real number, representing the utility of $h$.
\end{itemize}
\end{definition}


As for the deontic dimension, 
\emph{objective} ought-to-do's come from the optimal actions for an agent: to have seen to it that $\phi$ is taken to be an objective obligation of an agent at a given index iff $\phi$ is an effect of all the optimal actions for that agent at that index. The optimality of such actions is relative to a dominance ordering, and this ordering  depends on the value of the histories in those actions (provided by $\mathbf{Value}$). In order to present the semantics for formulas involving the ought-to-do operator, we therefore need some previous definitions. 

For $m\in M$ and $\beta\in Ags$, we define $\mathbf{State}_\beta^m=\left\{S\subseteq H_m; S=\bigcap_{\alpha \in Ags-\{\beta\}} s(\alpha), \mbox{ where }s\in \mathbf{Select}^m\right\}$. For $\alpha\in Ags$ and $m_*\in M$, we first define
a general ordering $\leq$ on $\mathcal{P}(H_{m_*})$ such that for $X, Y\subseteq H_{m_*}$, $
X\leq Y \textnormal{ iff } \mathbf{Value}(h) \leq \mathbf{Value}(h') \textnormal{ for every } h\in X, h'\in Y$. The objective dominance ordering $\preceq$ is defined such that for $L, L'\in \mathbf{Choice}_\alpha^{m_*}$, $
L\preceq L' \textnormal{ iff }  \mbox{for each } S\in \mathbf{State}_\alpha^{m_*}, L\cap S \leq L'\cap S.$ The optimal set of actions is the set $\mathbf{Optimal}^{m_*}_\alpha:=\{L \in \mathbf{Choice}^{m_*}_\alpha ; \textnormal{there is no } L' \in \mathbf{Choice}^{m_*}_\alpha \textnormal{such that }  L\prec L'\}.$    

As for \emph{subjective} ought-to-do's, they involve a dominance ordering as well, but one different to the one for objective ought-to-do's. To define this subjective dominance ordering, \cite{JANANDI} introduces a new semantic concept known as \emph{epistemic clusters}, which are nothing more than a given action's epistemic equivalents in indices that are indistinguishable to the one of evaluation. Formally, we have that for  $\alpha\in Ags$, $m_*, m\in M$, and $L\subseteq H_{m_*}$, $L$'s \emph{epistemic cluster} at $m$ is the set $[L]^m_\alpha:=\{h\in H_m ; \exists h_*\in L \ \textnormal{ s.t. }\ \langle m_*,h_*\rangle \sim_\alpha \langle m,h\rangle \}.$ As a convention, we write $m \sim_\alpha m'$ if there exist $h\in H_m$, $h'\in H_{m'}$ such that $\langle m,h\rangle \sim_\alpha \langle m',h'\rangle$. A subjective dominance ordering $\preceq_s$ is then defined on $\mathbf{Choice}^{m_*}_\alpha$ by the following rule: for $L, L'\subseteq H_{m_*}$, $L\preceq_s L'$ iff $\mbox{for each } m$ such that $m_*\sim_\alpha m$, $\mbox{for each } S\in \mathbf{State}_\alpha^{m}, [L]^m_\alpha\cap S \leq [L']^m_\alpha\cap S.$ Just as in the case of objective ought-to-do's, this ordering allows us to think about a subjectively optimal set of actions $\mathbf{S-optimal}^{m_*}_\alpha:=\{L \in \mathbf{Choice}^{m_*}_\alpha ; \textnormal{ there is no } L' \in \mathbf{Choice}^{m_*}_\alpha \textnormal{ s. t. }  L\prec_s L'\},$ where we write $L\prec_s L'$ iff $L\preceq_s L'$ and $L'\npreceq_s L$. Analogous to what we mentioned regarding objective ought-to-do's, the idea is that to have seen to it that $\phi$ is a subjective obligation of an agent at a given index iff it is an effect of all the subjectively optimal actions---and their epistemic equivalents---for that agent at that index.  

As is customary, the models and the semantics for the formulas are defined by adding a valuation function to the frames of Definition \ref{frames}: 
\begin{definition}
\label{models KCSTIT}
An \emph{eaubt}-model $\mathcal {M}$ consists of the tuple that results from adding a valuation function $\mathcal{V}$ to a \emph{eaubt}-frame, where $ \mathcal{V}: P\to 
    2^{M \times H}$ assigns to each atomic proposition a set of moment-history pairs.
    Relative to a model $\mathcal{M}$, the semantics for the formulas of $\mathcal {L}_{\mathtt{KOBO}}$ is defined recursively by the following truth conditions, evaluated at a given index $\langle m,h \rangle$: 
    \small
\[ \begin{array}{lll}
\mathcal{M},\langle m,h \rangle \models p & \mbox{ iff } & \mathcal{M},\langle m,h \rangle \in \mathcal{V}(p) \\

\mathcal{M},\langle m,h \rangle \models \neg \phi & \mbox{ iff } & \mathcal{M},\langle m,h \rangle \not\models \phi \\

\mathcal{M},\langle m,h \rangle \models \phi \wedge \psi & \mbox{ iff } & \mathcal{M},\langle m,h \rangle \models \phi \mbox{ and } \mathcal{M},\langle m,h \rangle \models \psi \\

\mathcal{M},\langle m,h \rangle \models \square \phi &
\mbox{ iff } & \mbox{for each } h'\in H_m, \mathcal{M},\langle m,h' \rangle \models \phi \\

\mathcal{M},\langle m,h \rangle \models [\alpha]
\phi & \mbox{ iff } & \mbox{for each } h'\in \mathbf{Choice}^m_\alpha(h), \mathcal{M},\langle m, h'\rangle \models \phi\\

\mathcal{M},\langle m,h \rangle \models K_{\alpha} \phi &
\mbox{ iff } & \mbox{for each } \langle m',h'\rangle \mbox{ s.t. }  \langle m,h \rangle \sim_{\alpha}\langle m',h' \rangle, \mathcal{M},\langle m',h' \rangle \models \phi\\


\mathcal{M},\langle m,h \rangle \models \odot[\alpha] \phi &
\mbox{ iff } & \mbox{for each } L\in \mathbf{Optimal} ^{m}_\alpha,  h'\in L \mbox{ implies that } \mathcal{M},\langle m,h' \rangle \models \varphi\\

\mathcal{M},\langle m,h \rangle \models \odot_{\mathcal{S}}[\alpha] \varphi & \mbox{ iff }  & \mbox{for each } L\in \mathbf{S-optimal} ^{m}_\alpha, \mbox{ for each } m' \mbox{ s.t. } m\sim_\alpha m', h'\in [L]^{m'}_\alpha \mbox{ implies that}\\&& 
\mathcal{M},\langle m,h' \rangle \models \varphi.
\end{array} \]
\normalsize
Satisfiability, validity on a frame, and general validity are defined as usual. We write $\|\phi\|$ to refer to the set $\{\langle m,h\rangle \in M\times H ;\mathcal{M},\langle m, h\rangle \models \phi\}$.

\end{definition}

\section{Introducing Beliefs}
\label{believeyoume}

The logic presented in the previous section offers many benefits for addressing complex interactive situations. The common thread among these situations is that agentive knowledge is taken into consideration when deciding whether an agent is responsible for having brought about some circumstance (see Horty's coin-flip puzzles in \cite{horty2019epistemic} and \cite{abarca2019logic}). However, we want to enhance the analysis by accounting for agents' belief-systems. As mentioned in the introduction, the beliefs that an agent has at a given index serve as justifications or explanations for a particular choice of action of said agent at said index.

In this work we adapt the arguments of  \cite{baltag2008probabilistic} and formalize a notion of \emph{full belief}. To clarify, an agent's full belief in the truth of a proposition means that the agent assigns probability 1 to the set of indices where the proposition is true. However, the typical logic of probabilistic full belief does not involve classical probability. The reason is that it is well known that classical measures yield problems for conditional beliefs and for belief revision. In classical-probability settings, conditioning on events with measure 0 is not defined and therefore ``it is unclear how to proceed if an agent learns something to which she initially assigned
probability 0'' (\cite{halpern2010lexicographic}, see also \cite{gkikas2015stable}, \cite{van1995fine}, and \cite{boutilier1995revision}). Since we want to incorporate to stit theory a paradigm of belief  that allows for revision, we follow the method of \cite{baltag2008probabilistic} and use \emph{conditional probability} as primitive.\footnote{There have been various attempts to deal with the problem of conditioning on events
of measure 0. The best known methods  involve (1) conditional-probability spaces (\cite{popper1968logic}, \cite{renyi1955new},  \cite{de1936probabilites}, \cite{van1995fine}, \cite{baltag2008probabilistic}, \cite{gkikas2015stable}), (2) nonstandard probability spaces (\cite{robinson1973function}, \cite{hammond1994elementary}, \cite{lehmann1992does}), where events with infinitesimally small probability may
still be learned or observed, and (3) lexicographic probability systems (\cite{halpern2010lexicographic}), which use sequences of probability measures with a descending order of importance.} Therefore, we build conditional-probability spaces upon the branching-time structures from Definition \ref{models KCSTIT}. To simplify the terminology---and just as done in  \cite{baltag2008probabilistic}---we focus on finite discrete structures, for which every subset is measurable with respect to special two-place probability functions mapping pairs of subsets to values in $[0,1]$. These functions underlie the semantics for formulas of conditional belief $B_\alpha^\psi \phi$, which are meant to be read as `after learning $\psi$, agent $\alpha$ believes that $\phi$ was the case (before the learning).'


\begin{definition}[Syntax with conditional-belief]
\label{syntax2}
The grammar for the formal language $\mathcal L_{\textsf{KOB}}$ (an extension of $\mathcal L_{\textsf{KO}}$) is given by: $\phi :=  p \mid \neg \phi \mid \phi \wedge \psi \mid \square \phi \mid [\alpha] \phi \mid K_\alpha \phi \mid  B_\alpha^\psi\phi \mid \odot [\alpha] \phi \mid \odot_{\mathcal{S}}[\alpha] \phi.$
\end{definition}

\begin{definition}[Epistemic act-utilitarian discrete-conditional-probability branching-time frames]
\label{models KB}
A \textbf{finite} \emph{epistemic act-utilitarian discrete-conditional-probability branching-time} frame (\emph{dcpbt}-frame for short) is a tuple $\langle M,\sqsubset,\mathbf{\mathbf{Choice}}, \{\sim_\alpha\}_{\alpha\in Ags}, \{\mu_\alpha\}_{\alpha\in Ags},  \mathbf{Value} \rangle$ such
that $M,\sqsubset,\mathbf{\mathbf{Choice}}, \{\sim_\alpha\}_{\alpha\in Ags}$, and $\mathbf{Value}$ are like in Definition \ref{frames}, and for every $\alpha\in Ags$, $\mu_\alpha:\mathcal{P}(M\times H)\times \mathcal{P}(M\times H)\to [0,1]$ is such that (a) for each $B\subseteq M\times H$, $\mu_\alpha^B:=\mu_\alpha(\cdot | B)$ is either a constant function with value 1 or a classical-probability function on $M\times H$, and (b) for every $A, B, C \subseteq M\times H$, $\mu_\alpha(A\cap B | C)=\mu_\alpha(A|B\cap C)\cdot\mu_\alpha(B|C)$.\footnote{This means that $\mu_\alpha$ is a two-place probability function that meets the following three axioms: (1) $\mu_\alpha(A | A)=1$ for every $A\subseteq M\times H$, (2) if $A\cap B=\emptyset$, $C\neq\emptyset$, and $\mu_\alpha^C$ is not constant 1, then $\mu_\alpha(A\cup B | C)=\mu_\alpha(A|C)+\mu_\alpha(B|C)$, and (3) $\mu_\alpha(A\cap B | C)=\mu_\alpha(A|B\cap C)\cdot\mu_\alpha(B|C)$. These three axioms are called the Popper-R\'enyi axioms for conditional-probability spaces, and they were first introduced in \cite{renyi1955new}. 
Observe that if $\mu_\alpha(B|M\times H)=0$, then this condition does not prevent $\mu_\alpha (\cdot|B)$ from being defined, and this is the quality of the theory of conditional-probability spaces that allows for conditioning on events with measure 0. We illustrate this property and its implications for conditional belief with the example included in Section \ref{giventofly}.}

A \emph{dcpbt}-model $\mathcal{M}$ results from adding a valuation function $\mathcal{V}$ to a \emph{dcpbt}-frame, and the semantics for the formulas of $\mathcal {L}_{\textsf{KOB}}$ over such a model is defined recursively as in Definition \ref{models KCSTIT}, with the following additional clause: $\mathcal{M},\langle m,h \rangle \models B_\alpha^\psi\phi$ iff $\mu_\alpha\left(\|\phi\| \ | \ \|\psi\| \cap \pi_\alpha[\langle m, h\rangle]\right)=1.$
\end{definition}

\begin{remark}\label{lolipop} In what follows, we will refer to any function $q:\mathcal{P}(M\times H)\times \mathcal{P}(M\times H)\to [0,1]$ that meets the requirements (a) and (b) of Definition \ref{models KB} as a \emph{vf}-function. Therefore, if $p:M\times H\to [0,1]$ is a classical-probability function, then the function $p_c:(M\times H)\times (M\times H)\to [0,1]$, defined by the rules $p_c(A|B)=\frac{p(A\cap B)}{P(B)}$ if $p(B)>0$ and $p_c(A|B)=1$ if $p(B)=0$, is a \emph{vf}-function (see \cite{van1995fine} and \cite[Chapter 3]{gkikas2015stable}). This means that probability theory's typical definition of conditional probability in terms of a classical-probability function is a special instance of a \emph{vf}-function. 
\end{remark}

Endowed with semantics for formulas involving conditional belief, we take plain (unconditional) \emph{belief} to be represented by beliefs conditional on a tautology. Therefore, in what follows we write $B_\alpha\phi$ to denote $B_\alpha^\top\phi$. With these formulas and the models they are evaluated on, we have extended the epistemic stit logic of Section \ref{shit} into a system that deals with both \emph{knowledge} and \emph{belief}. As pointed out in \cite{baltag2006logic}, \cite{baltag2006conditional}, and \cite{baltag2008probabilistic}, the truth conditions in Definition \ref{models KB} yield a logic for which the knowledge operators validate the $\mathbf{S5}$ schemata, the conditional-belief operators validate the $\mathbf{K}$ schema, and the following interaction schemata are valid: $K_\alpha\phi\to B_\alpha^\psi\phi$ (\emph{Persistence of knowledge}); $B_\alpha^\psi\phi\to K_\alpha B_\alpha^\psi\phi$ and $\lnot B_\alpha^\psi\phi\to K_\alpha\lnot B_\alpha^\psi\phi$ (\emph{Full introspection of belief}). Additionally, the following axioms that regard revision are also valid: $B_\alpha^\phi\phi$ (\emph{Hypotheses are accepted}); $B_\alpha^\psi\phi \to (B_\alpha^{\phi\land \psi}\theta\leftrightarrow B_\alpha^\psi \theta)$ and  $\lnot B_\alpha^\psi \lnot \phi \to (B_\alpha^{\phi\land \psi}\theta\leftrightarrow B_\alpha^\psi (\phi\to \theta))$ (\emph{Minimality of revision}); $\phi\to\lnot B_\alpha^\phi\bot$ (\emph{Weak consistency of belief}).\footnote{Observe that the semantics for conditional belief implies that an agent's belief of $\phi$ given $\psi$ is relative to the agent's epistemic state. In other words, we take it that conditional beliefs depend on the \emph{information} available to the agent.}

Why account for belief revision? Well, in \cite{halpern2010lexicographic} Halpern argues that belief revision plays a critical role in the analysis of strategic reasoning in extensive-form games, and since the stit semantics for agency over branching-time structures can adopt most ideas of the theory of extensive-form games (see \cite[Chapter 7]{Horty2001}, \cite{duijf2016representing}, and \cite{abarca2019stit}), we believe that stit theory should also benefit from an account of conditional belief and of belief revision. There is a common idea that if a formalization of the temporal evolution of a game incorporates the assumption that players can change their beliefs about the game as it progresses due to some flow of information---for instance by drawing conclusions from opponent's moves---then the analysis of interactive situations becomes much richer (\cite{board2004dynamic}).\footnote{One can easily associate a belief-revision Kripke-structure to an extensive-form game to evaluate formulas that reflect agents' belief changes as the game ``progresses'' (see \cite{board2004dynamic} for an illustration of such an association in the analysis of \emph{backward induction} in extensive-form games). The typical way to do so is to think of full strategy profiles---functions that associate each node to a player's move (or strategy) at that node---as possible worlds, so that the formulas evaluated on these worlds ``[...] describe the way the game is actually played, and they provide a set
of counterfactuals for evaluating the payoffs if the action taken at any node deviates from the specified action.'' The evaluation of formulas at full strategy profiles is reminiscent of stit theory's use of histories as part of the indices of evaluation, but in extensive-form games one also considers profiles that cannot be realized in the structure of the game: regardless of whether the nodes can be reached in a \emph{play} or not, the functions that serve as possible worlds map each node to a move (or strategy) that can be played at that node. It is this feature of the belief-revision structure associated to an extensive-form game that makes conditional-belief modalities very useful. With the syntax of conditional-belief, we can represent the beliefs of an agent at a node that was actually reached in a play, by conditioning on  formulas that ensure that such a node was reached. Although this feature of the language and the semantics of conditional belief could be very well exploited in the context of \emph{strategic} stit theory (\cite[Chapter 7]{Horty2001}), here we propose that it is also relevant even in instantaneous-stit theory, namely due to agents' uncertainty across the set of indices. If an agent had uncertainty about the index of evaluation it found itself at (because of some lack of information), then the conditional-belief modalities allow us to reason syntactically about the counterfactual situation in which the agent's learning of such information would change its state of uncertainty.} 

In stit theory, branching-time structures represent an exhaustive set of possibilities for temporal evolution according to multi-agent interaction, and a treatment of belief change allows for a description of how agents could have constrained those possibilities if they had learned information regarding past choices about which they are uncertain. For instance, in the example that we mentioned in the introduction, the doctor did not know that the patient had eaten before being admitted to the hospital, and moreover the doctor \emph{learned}---from the mistaken chart---that the patient had not eaten. In principle, if the doctor had learned that the patient had in fact eaten, then her beliefs should have been different. In other words, she would \emph{revise} her beliefs after learning that the patient had eaten before being admitted to the hospital. By incorporating conditional-belief modalities into stit theory, we can analyze both syntactically and semantically the different ways in which doxastic states can change according to the learning of information. In this way, we can formalize further the justification of choices of action that have moral consequences.

\section{Introducing Doxastic Oughts}
\label{giventofly}

In keeping with EGT, we argue that agents can be seen as having a doxastic sense of what they ought to do in interactive situations, and that this sense can be traced back both to their beliefs regarding the index at which they are and to the utilities of the histories in their available actions. Rather than delving into the analyses of rationality and rational choice in terms of best responses (see \cite{sep-decision-theory} and \cite{sep-epistemic-game} for surveys of such analyses), we use the concepts of \emph{utility} and \emph{belief} in order to characterize a doxastic sense of ought-to-do in stit theory. Here, utility and belief underlie the process by which the beliefs of an agent explain whether that agent was justified in making a particular choice of action. Again, this extension of deontic stit is important to treat the kind of problems in excusability and responsibility attribution that stit theory deals with (\cite{lorini2014logical}), as we will illustrate by analyzing the example presented in the introduction. 

Our interpretation of belief rests upon probabilities assigned to the alternative histories of branching-time structures. Inspired by the customary treatment of decision rules under \emph{uncertainty} (or \emph{risk}) from decision theory,\footnote{\label{athens}It is convenient to remark about subtle differences in definitions for overlapping concepts. Decision theorists typically distinguish between \emph{choice under uncertainty}, for which decision makers do not know the outcome of a decision they engage in, and \emph{choice under risk}, for which decision makers have probabilistic information regarding the outcomes. In decision theory, when agents have such probabilistic information for choices under risk, it means that they have \emph{objective} information regarding the outcomes. In other words, the probabilistic information is not meant to embody an agent's subjective beliefs regarding the outcomes. When \emph{subjective} probabilities are introduced, like the ones we deal with here, the general view is to treat these cases as involving choice under uncertainty (see \cite{peterson2017introduction}, \cite{sharpe2018risk}). } we identify an agent's doxastic sense of ought-to-do with the effects of actions that maximize \emph{expected (deontic) utility}. 

Expected utility theory has somewhat settled interpretations for the components of interactive decision making (see \cite{Savage1954}, \cite{jeffrey1965ethics}, and \cite{karni2014axiomatic}). Typically, the utilities of outcomes quantify agents' \emph{preferences}, and the probabilities assigned are seen either as \emph{objective} measures for the frequency with which sets of outcomes---the \emph{events}---ensue or as \emph{subjective} representations of agentive belief (see Footnote \ref{athens}). As mentioned before, in the context of deontic stit based on act utilitarianism we interpret the value of a  history---$\mathbf{Value}(h)$---as its deontic utility for the whole group of agents, with no specific interpretation for the word `utility.' This means that we do not identify the deontic utility of a history with agentive \emph{preference} but rather allow for the assignment of values to ``accommodate a variety of different approaches'' (see Section 2.2 \cite[Chapter 3]{Horty2001}). The notion of deontic utility itself is taken as primitive in act utilitarian stit, and it is a general notion that applies to the whole set of agents---not only to individual ones. Therefore, it may be thought of---but not necessarily so---as the ``total utility of the set of agents in that history, their average utility, or perhaps some distribution-sensitive aggregation of the utilities of these individual agents'' (\cite{Horty2001}, p. 38). As for the probabilities, it must be clear by now that we interpret them as a measure of the agents' individual doxastic state. 



\begin{definition}

Let $\mathcal{M}$ be a \emph{dcpbt}-model. Let $m\in M$, $h\in H_m$, and $\alpha\in Ags$. Let $L\in \mathbf{Choice}_\alpha^m$. We define $\alpha$'s \emph{expected deontic utility} of $L$ at $\langle m,h \rangle$---denoted by $EU_\alpha^{\langle m,h\rangle}(L)$---as the value given by the following formula: $EU_\alpha^{\langle m,h\rangle}(L):=\sum\limits_{m'\sim_\alpha m, h'\in [L]_\alpha^{m'}}\mu_\alpha\left( \{h'\} \ | \  \pi_\alpha[\langle m',h' \rangle]\right)\cdot \mathbf{Value}(h')$.
\end{definition}

\begin{remark}
This means that we calculate $\alpha$'s expected deontic utility for one of its available actions $L$ at $m$ by summing the utilities of all the histories lying in the \emph{epistemic clusters} of $L$, weighted by the probabilities that $\alpha$ assigns to those histories, conditional on $\alpha$'s information set at the index where $\alpha$ is. 
Observe that for every $m\in M$ and $L\in \mathbf{Choice}_\alpha^m$, we have that $EU_\alpha^{\langle m,h\rangle}(L)=EU_\alpha^{\langle m,h'\rangle}(L)$ for every $h,h'\in H_m$.
\end{remark}

Our notion of expected deontic utility can be seen as a stit version of EGT's interpretation of an agent's expected utility for a given strategy, conditional on that agent's information. According to \cite{sep-epistemic-game}, in EGT an agent's expected utility for one of its strategies is calculated with respect to so-called \emph{conjectures}. An agent's conjecture at a given world is a probability distribution on the set of all strategy profiles involving the \emph{other} agents. Such a distribution is typically based on probabilities conditional either on the agent's strategy-choice at the world of evaluation or on the agent's information set at said world (\cite{sep-epistemic-game}). Our framework differs from EGT's in three essential points: (1) we do not have individual utilities for each outcome; (2) in our structures, each \emph{possible world} (which we refer to as an \emph{index}) has a deontic utility, whereas in EGT only outcomes---or full strategy profiles---have utilities; and (3) while EGT regards information sets as subsets of the available strategies, we do not impose this condition---in fact, the semantic condition $(\mathtt{OAC})$ that we adopt yields that information sets are unions of choice cells. 

Since \emph{dcpbt}-models are finite, we have that for every $m\in M$ and $h\in H_m$, the set $\{EU_\alpha^{\langle m,h\rangle}(L); L\in \mathbf{Choice}_\alpha^m\}$ has a maximum. Therefore, there are actions that maximize $\alpha$'s expected deontic utility at every index, namely the ones whose expected deontic utility is the same as said maximum. We denote by $\mathbf{EU}_\alpha^{\langle m,h\rangle}$ the set of  actions that maximize $\alpha$'s expected deontic utility at ${\langle m,h\rangle}$.\footnote{Formally, our definition of expected deontic utility is in fact an instance of probability theory's \emph{conditional expectation with respect to an event}---albeit in the setting where conditional probability is primitive. In this case, the so-called \emph{event} is the information set of a given agent at the index of evaluation. Thus, if $L$ is an available action for an agent at index $\langle m,h\rangle$, and if $E$ denotes the expected value of the random variable $\mathbf{Value}$ with respect to $\mu_\alpha(\cdot | H)$ (where we recall that $H$ is the set of all histories), then $EU_\alpha^{\langle m,h\rangle}(L)=E(\mathbf{Value}|\pi_\alpha[\langle m,h'\rangle])$ for any $h'\in L$.} 



\begin{definition}[Full syntax]
\label{syntax3}
The grammar for the formal language $\mathcal L_{\textsf{KOBO}}$ (an extension of $\mathcal L_{\textsf{KOB}}$) is given by: $\phi :=  p \mid \neg \phi \mid \phi \wedge \psi \mid \square \phi \mid [\alpha] \phi \mid K_\alpha \phi \mid  B_\alpha^\psi\phi \mid \odot [\alpha] \phi \mid \odot_{\mathcal{S}}[\alpha] \phi \mid \odot_{\mathcal{B}}[\alpha] \phi.$
\end{definition}

We use the same \emph{dcpbt}-models from Definition \ref{models KB} to evaluate the formulas of $\mathcal L_{\textsf{KOBO}}$. The truth conditions are the same as before, with the following additional clause: 
$\mathcal{M},\langle m,h \rangle \models \odot_{\mathcal{B}}[\alpha] \phi$
iff for each $L \in \mathbf{EU}_\alpha^{\langle m,h\rangle}$  we have that $[L]^{m'}_\alpha \subseteq \|\phi\|$  for all  $m'$ such that $m\sim_\alpha m'.$ In other words, at a given index it was the case that an agent doxastically ought to have seen to it that $\phi$ iff $\phi$ is an effect both of all the actions that maximized the agent's expected deontic utility at said index and of the epistemic equivalents of these actions.
\subsection{Example}

In order to illustrate the reach of the semantics introduced above, we present a formal analysis of the example in the introduction, using \emph{dcpbt} models. 

\begin{figure}[htb!]
\begin{minipage}[c]{.9\linewidth}
				\centering
\resizebox{220pt}{!}{%
\begin{tikzpicture}[level distance=2cm,
level 1/.style={sibling distance=6cm},
level 2/.style={sibling distance=4.2cm},
level 3/.style={sibling distance=1.5cm}
]

\node {} [grow=up]
	child{node [matrix, matrix of nodes, ampersand replacement=\&, label=left:$m_1$, label=below right:$Choice_{patient}^{m_1}$] (matrixi) {
	\node[matrix node, label=below:$E_1$] (m11) {}; \& \node[matrix node, label=below:$E_2$] (m12) {}; \\ }  
		child[noline]{node [matrix, matrix of nodes, ampersand replacement=\&, label=right:$m_3$, label=below right:$Choice_{doctor}^{m_3}$]{\node[matrix node, label=below:$L_3$] (m31) {};  \& \node[matrix node, label=below:$L_4$] (m32) {}; \\}
			child{node (h5) {$h_4$}
				}
            child{node (h4) {$h_3$}}
			}
		child[noline]{node [matrix, matrix of nodes, ampersand replacement=\&, label=left:$m_2$, label=below left:$Choice_{doctor}^{m_2}$]{\node[matrix node, label=below: $L_1$] (m21) {};  \& \node[matrix node, label=below:$L_2$] (m22) {}; \\}
			child{node (h2) {$h_2$}
				}
           	child{node (h1) {$h_1$}
				}
			}
		};

\draw (m11.center) -- (m22);
\draw (m12.center) -- (m31);
\draw (m21.center) -- (h1) node [pos=.7,draw=none, label=above left:\footnotesize$1$] (lh1) {};
\draw (m21.center) -- (h1) node [pos=.5,draw=none, label=above left:\footnotesize$a\land r$]  {};
\draw (m21.center) -- (h1) node [pos=.3,draw=none, label=above left:\footnotesize$\lnot e$] {};
\draw (m22.center) -- (h2) node [pos=.7,draw=none, label=above right:\footnotesize$0$] (lh2) {};
\draw (m22.center) -- (h2) node [pos=.5,draw=none, label=above right:\footnotesize$\lnot a\land \lnot r$]  {};
\draw (m22.center) -- (h2) node [pos=.3,draw=none, label=above right:\footnotesize$\lnot e$]  {};
\draw (m31.center) -- (h4) node [pos=.7,draw=none, label=above left:\footnotesize$-1$] (lh4) {};
\draw (m31.center) -- (h4) node [pos=.5,draw=none, label=above left:\footnotesize$a\land d$]  {};
\draw (m31.center) -- (h4) node [pos=.3,draw=none, label=above left:\footnotesize$e$]  {};

\draw (m32.center) -- (h5) node [pos=.7,draw=none, label=above right:\footnotesize$0$] (lh5) {};
\draw (m32.center) -- (h5) node [pos=.5,draw=none, label=above right:\footnotesize$\lnot a\land\lnot r$] {};
\draw (m32.center) -- (h5) node [pos=.3,draw=none, label=above right:\footnotesize$e$] {};

\draw[dashed,blue, bend right] (lh1) to (lh4) {};
\draw[dashed,blue, bend right] (lh2) to (lh5) {};

\end{tikzpicture}
}
\captionof{figure}{Example from \emph{The Verdict}}
\label{fig:fig1}
\end{minipage}
\end{figure}
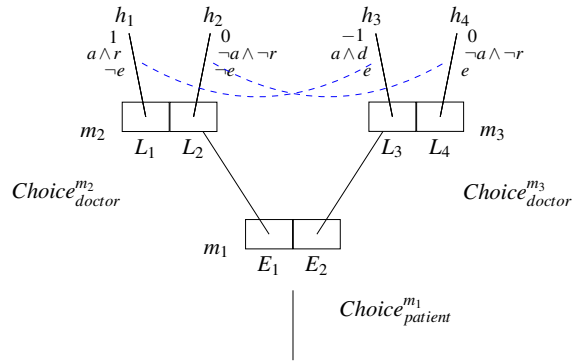

\begin{example}
\label{ex1}
 Let $Ags=\{patient, doctor\}$. Let $M$ and $\sqsubset$ be defined so as to be represented by the diagram in Figure \ref{fig:fig1}. We have three moments ($m_1-m_{3}$) and four histories ($h_1-h_{4}$). These histories represent the different possibilities in which time may have evolved according to the actions available both to $patient$ and $doctor$. The actions available to $patient$ at moment $m_1$ are $E_1$, which we interpret as the action of refusing to eat, and $E_2$, which we interpret as the action of eating. It is according to such actions that time progressed either into moment $m_2$ or into moment $m_3$. At both these moments, $doctor$ chose from her available actions and executed one of them. At moment $m_2$, the actions available to $doctor$ are $L_1$, which we interpret as the action of supplying anesthetics, and $L_2$, which we interpret as the action of refusing to supply anesthetics. Similarly, at moment $m_3$ the actions available to $doctor$ are $L_3$, which we interpret as the action of supplying anesthetics, and $L_4$, which we interpret as the action of refusing to supply anesthetics. 

We model the utilities of the outcomes according to the statement of the example. Therefore, history $h_1$, where at $m_1$ $patient$ refused to eat and at $m_2$ $doctor$ supplied the anesthetics, gets the highest utility---$\mathbf{Value}(h_1)=1$---due to the fact that such a history represents the situation in which the $patient$ got ready for the surgery without any trouble. Histories $h_2$ and $h_4$ get a neutral utility of 0: both of them imply that $doctor$ refused to supply anesthetics and thus the patient is not ready for the surgery. History $h_3$, on the other hand, gets a negative utility of $-1$, since it implies that at $m_1$ $patient$ ate and at $m_3$ $doctor$ supplied the anesthetics, leading to $patient$'s death. As implied by the statement of the example, we take $h_3$ to be the \emph{actual} history.  

The epistemic-doxastic states and ought-to-do's that we focus on are those of $doctor$, since they illustrate important semantic properties of our logic. We represent $doctor$'s epistemic states with blue dashed lines, so that at both $m_2$ and $m_3$, along every history running through them, $doctor$ did not know whether the patient had eaten or not, but she knew which action she performed. The doxastic states of $doctor$ are represented by the \emph{conditional-probability} function $\mu_{doctor}$, given by the following rules. 
Let $p_{doctor}:\mathcal{P}(M\times H) \to [0,1]$ be a discrete classical-probability function such that $p_{doctor}(\langle m_i, h_j \rangle)=\frac{.9}{4}$ for $i,j\in\{1,2\}$, $p_{doctor}(\langle m_1, h_i \rangle)=\frac{.1}{4}$ for $i\in\{3,4\}$, and $p_{doctor}(\langle m_3, h_i \rangle)=\frac{.1}{4}$ for $i\in\{3,4\}$. We then define $\mu_\alpha$ so that $\mu_{doctor}(A|B)=\frac{p_{doctor}(A\cap B)}{p_{doctor}(B)}.$
\end{example}

Let $e$ denote the atomic proposition `the patient has eaten,' let $a$ denote the atomic proposition `anesthetics are supplied to the patient,' let $r$ denote the atomic proposition `the patient is ready for surgery,' and let $d$ denote the atomic proposition `the patient will die.' According to Definition \ref{models KB}, these atomic propositions and the formulas that are recursively built with them can be taken as true or false depending on the index of evaluation. For instance, we model the example so that at index $\langle m_2, h_1\rangle$ it was the case that $patient$ ate, that $doctor$ supplied the anesthetics, and that $patient$ became ready for surgery. ($\mathcal{M},\langle m_2,h_1 \rangle \models e\land a \land r$). 

As for the evaluation of formulas involving the basic stit-theory operators, observe that some instances of it are $\mathcal{M},\langle m_2,h_1 \rangle \models \square e$ (\emph{at this index it was the case that it was settled that the patient did not eat}), $\mathcal{M},\langle m_3,h_3 \rangle \models [doctor] d$  (\emph{at this index it was the case that the doctor saw to it that the patient died}). As for formulas involving the epistemic-doxastic operators, we have that for $i\in \{2,3\}$ and $j\in \{1,2,3,4\}$, $\mathcal{M},\langle m_i,h_j \rangle \models \lnot K_{doctor}e \land \lnot K_{doctor}\lnot e$ (\emph{at said indices it was the case that the doctor did not know whether the patient had eaten or not}) and $\mathcal{M},\langle m_i,h_j \rangle \models K_{doctor}[doctor]a \lor K_{doctor}[doctor]\lnot a$ (\emph{at said indices it was the case that the doctor knew whether she was supplying the anesthetics or not}). Moreover, we have that $\mathcal{M},\langle m_3,h_3 \rangle \models \lnot B_{doctor}\lnot e$ (\emph{at the actual index it was not the case that the doctor \emph{fully and unconditionally} believed that the patient had not eaten}), that $\mathcal{M},\langle m_3,h_3 \rangle \models B_{doctor}^{e}d$ (\emph{at the actual index it was the case that if the doctor had learned that the patient had in fact eaten, then she would have fully believed that the patient would die}), and that  $\mathcal{M},\langle m_3,h_3 \rangle \models B_{doctor}^{e}[doctor]d$ (\emph{at the actual index it was the case that if the doctor had learned that the patient had in fact eaten, then she would have fully believed that she would kill the patient}). 

As for formulas involving the deontic operators, we first observe that \begin{itemize}
    \item $\mathbf{Optimal}_{doctor}^{m_2}=\{L_1\}$, $\mathbf{Optimal}_{doctor}^{m_3}=\{L_4\}$.
    \item $\mathbf{S-Optimal}_{doctor}^{m_2}=\{L_1, L_2\}$, $\mathbf{S-Optimal}_{doctor}^{m_3}=\{L_3, L_4\}$.  
    \item $\mathbf{EU}_{doctor}^{\langle m_2, h_i \rangle}=\{L_1\}$ ($i\in\{1,2\}$), $\mathbf{EU}_{doctor}^{\langle m_3, h_i \rangle}=\{L_3\}$ ($i\in\{3,4\}$).
\end{itemize}   

Therefore, for instance we have that $\mathcal{M},\langle m_2,h_i \rangle \models \odot[doctor]a\land \lnot K_{doctor}\odot[doctor]a$ ($i\in\{1,2\}$) (\emph{at moment $m_2$, along any history running through it, it was the case that the doctor objectively ought to have supplied the anesthetics, but the doctor did not know that for sure}), that $\mathcal{M},\langle m_3,h_i \rangle \models \odot[doctor]\lnot a\land \lnot K_{doctor}\odot[doctor]\lnot a$ ($i\in\{3,4\}$) (\emph{at moment $m_3$, along any history running through it, it was the case that the doctor objectively ought to have refrained from supplying the anesthetics, although the doctor did not know that}), that  $\mathcal{M},\langle m_i,h_j \rangle \models \lnot \odot_{\mathcal{S}}[doctor]a \land \lnot \odot_{\mathcal{S}}[doctor]\lnot a$ ($i\in\{1,2\}$ and $j\in\{1,2,3,4\}$) (\emph{at no index it was the case that the doctor either subjectively ought to have supplied the anesthetics or subjectively ought to have refrained from supplying them}), that $\mathcal{M},\langle m_i,h_j \rangle \models \odot_{\mathcal{B}}[doctor]a \land  K_{doctor}\odot_{\mathcal{B}}[doctor] a$ ($i\in\{2,3\}$ and $j\in\{1,2,3,4\}$) (\emph{at every index it was the case that the doctor doxastically ought to have supplied the anesthetics and that she knew that}). Observe that we could model the fact that $patient$'s information chart was mistaken by the fact that $\mathcal{M},\langle m_3,h_3 \rangle \models B_{doctor}^{\lnot e}(\lnot e \land \odot[doctor]a)$ (\emph{at the actual index it was the case that if the doctor had learned that the patient had not eaten (as she did by the mistake in the chart), then she would have \emph{fully} believed that the patient had not eaten and that she objectively ought to have supplied the anesthetics}.) Coupled with the fact that $doctor$ did not know that $patient$ had in fact eaten, the satisfaction of these last two formulas at the actual index should in principle provide a good reason for \emph{excusing} $doctor$ of actually having caused $patient$'s death.

\section{Axiomatization and Logic Properties}
\label{axiomatique}

Since one of the main contributions of the present paper is the introduction of doxastic oughts, we review some of the properties of the semantics for formulas involving the operator $\odot_{\mathcal{B}}[\alpha]$. First of all, we must say that this modal operator yields a $\mathbf{KD45}$ logic. Secondly, the doxastic sense of ought validates a version of Kant's imperative \emph{ought implies can}, as explained by the fact that the following formula is valid with respect to the class of \emph{dcpbt} models: $\odot_{\mathcal{B}}[\alpha ]\phi\to\Diamond K_\alpha \phi.$ We also have that if an agent doxastically ought to have seen to it that $\phi$, then the agent knows that this is settled---the formula $\odot_{\mathcal{B}}[\alpha ] \phi \to K_{\alpha} \square\odot_{\mathcal{B}}[\alpha ] \phi$ is valid as well. As for the interaction between this doxastic sense of ought and the objective/subjective ought-to-do's, we have that $\not\models  \odot_{\mathcal{B}}[\alpha] \varphi \to  \odot[\alpha] \varphi$ and that $\not\models \odot[\alpha] \varphi \to  \odot_{\mathcal{B}}[\alpha] \varphi$ (as can be inferred from Example \ref{ex1}). Similarly, we have that $\not\models \odot[\alpha]_{\mathcal{B}} \varphi \to  \odot_{\mathcal{S}}[\alpha] \varphi$ and that $\not\models  \odot_{\mathcal{S}}[\alpha] \varphi \to  \odot_{\mathcal{B}}[\alpha] \varphi$. The first invalidity can be inferred from Example \ref{ex1}. The second one can be inferred from a variation of Example \ref{ex1} as follows: let $\mathbf{Value}(h_1)=2$, $\mathbf{Value}(h_2)=1$, $\mathbf{Value}(h_3)=0$, and $\mathbf{Value}(h_4)=0$; let $p_{doctor}$ be a discrete classical-probability function such that $p_{doctor}(\langle m_2, h_1 \rangle)=\frac{.1}{2}$, $p_{doctor}(\langle m_2, h_2 \rangle)=\frac{.9}{2}$,  $p_{doctor}(\langle m_3, h_3 \rangle)=\frac{.9}{2}$, $p_{doctor}(\langle m_3, h_4 \rangle)=\frac{.1}{2}$, and $p_{doctor}$ has constant value 0 on all other indices; let $\mu_\alpha$ be defined as in Example \ref{ex1}; then it is the case that $\mathcal{M},\langle m_i,h_j \rangle \models \odot_{\mathcal{S}}[doctor]a \land  \odot_{\mathcal{B}}[doctor]\lnot a$ ($i\in\{2,3\}$ and $j\in\{1,2,3,4\}$). 

In order to further the understanding of our logic, we present a sound, complete, and decidable proof system for it. 
\begin{definition}[Proof system]
\label{axiomsystemurakami}

Let $\Lambda$ be the proof system defined by the following axioms and rules of inference: 
\begin{itemize}
\item \emph{(Axioms)} All classical tautologies from propositional logic. The $\mathbf{S5}$ axiom schemata for $\square$, $[\alpha]$, $K_\alpha$. The following axioms and schemata for the interactions of formulas with the given operators:
\footnotesize
\[\begin{array}{ll}
\odot [\alpha ] (\phi\to \psi)\to (\odot [\alpha ] \phi \to \odot [\alpha ] \psi)& (A1)\\ 
\square \phi\to [\alpha ] \phi \land \odot [\alpha ] \phi& (A2)\\ 
\square\odot [\alpha ] \phi\lor  \square\lnot\odot [\alpha ] \phi&(A3)\\ 
\odot [\alpha ] \phi\to  \odot [\alpha ] ([\alpha ]\phi)&(A4)\\
\odot [\alpha ] \phi\to \Diamond  [\alpha ] \phi &(Oic)\\
\mbox{For $n\geq 1$ and pairwise different $\alpha_1,\dots,\alpha_n$},& \\ \bigwedge_{1\leq k\leq n}  \Diamond [\alpha_i ] \phi_i \to \Diamond\left(\bigwedge_{1\leq k\leq n}[\alpha_i ] \phi_i\right)& (IA)\\
K_\alpha \phi\to [\alpha ]\phi&(OAC)\\
\Diamond K_\alpha \phi \to K_\alpha \Diamond  \phi&(Unif-H)\\
\odot_{\mathcal{S}}[\alpha ] (\phi\to \psi)\to (\odot_{\mathcal{S}} [\alpha ] \phi \to \odot_{\mathcal{S}}[\alpha ] \psi)& (A5)\\
\odot_{\mathcal{S}} [\alpha ] \phi\to  \odot_{\mathcal{S}}[\alpha ] (K_\alpha \phi)&(A6)\\
K_\alpha \square \phi\to\odot_{\mathcal{S}}[\alpha ]\phi &(s.N)\\ \odot_{\mathcal{S}}[\alpha ]\phi\to\Diamond K_\alpha \phi & (s.Oic)\\
\odot_{\mathcal{S}}[\alpha ] \phi \to K_{\alpha} \square\odot_{\mathcal{S}}[\alpha ] \phi  &(s.Cl1)\\
\lnot\odot_{\mathcal{S}}[\alpha ] \phi \to K_{\alpha} \square\lnot\odot_{\mathcal{S}}[\alpha ] \phi  &(s.Cl2)\\
B_\alpha(\phi\to \theta)\to (B_\alpha\phi\to B_\alpha \theta)&(A7)\\
(\psi\leftrightarrow \phi)\to (B_\alpha\theta\leftrightarrow B_\alpha^\phi \theta)&(A8)\\
K_\alpha \phi\to B_\alpha\phi&(PK)\\
B_\alpha \phi\to K_\alpha B_\alpha\phi &(FIB1)\\ \lnot B_\alpha \phi\to K_\alpha \lnot B_\alpha\phi &(FIB2)\\ 
B_\alpha^\phi \phi&(HA)\\
B_\alpha \phi\to (B_\alpha^{\phi\land \psi}\theta\leftrightarrow  B_\alpha \theta) &(MBR1)\\
\lnot B_\alpha \lnot \phi\to (B_\alpha^{\phi\land \psi}\theta\leftrightarrow  B_\alpha(\phi\to \theta))&(MBR2)\\
\psi\to \lnot B_\alpha\bot&(WCon)\\
\odot_{\mathcal{B}}[\alpha ] (\phi\to \psi)\to (\odot_{\mathcal{B}} [\alpha ] \phi \to \odot_{\mathcal{B}}[\alpha ] \psi)& (A9)\\
\odot_{\mathcal{B}} [\alpha ] \phi\to  \odot_{\mathcal{B}}[\alpha ] (K_\alpha \phi)&(A10)\\
K_\alpha \square \phi\to\odot_{\mathcal{B}}[\alpha ]\phi&(d.N)\\
\odot_{\mathcal{B}}[\alpha ]\phi\to\Diamond K_\alpha \phi&(d.Oic)\\
\odot_{\mathcal{B}}[\alpha ] \phi \to K_{\alpha} \square\odot_{\mathcal{B}}[\alpha ] \phi  &(d.Cl1) \\
\lnot\odot_{\mathcal{B}}[\alpha ] \phi \to K_{\alpha} \square\lnot\odot_{\mathcal{B}}[\alpha ] \phi &(d.Cl2)
\end{array}\]
\item \textit{(Rules of inference)} \textit{Modus Ponens}, Substitution, and Necessitation for all modal operators.
\end{itemize}
\end{definition}

In the Appendix of this work, we discuss all these axioms and schemata, and we show that the proof system $\Lambda$ is sound and complete with respect to a class of models that are more general than the ones introduced in Definition \ref{models KB}. These models differ from \emph{dcpbt}-models in two main qualities: (1) following \cite{abarca2019logic}, they are multi-valued to the extent that instead of only one deontic value function, they have three: one for the objective ought-to-do's, one for the subjective ones, and one for the doxastic ones; and (2) they are Kripke-structures based on domains of possible worlds.\footnote{Nevertheless, one can adapt the correspondence between Kripke models and branching-time models from \cite{xu1994decidability} and \cite{abarca2019logic} to show that $\Lambda$ is sound and complete with respect to the class of \emph{multi-valued} \emph{dcpbt} models.} The proof of completeness in the Appendix also shows that $\Lambda$ is \emph{decidable}. This is a consequence of the logic's finite model property, which is shown through obtaining a finite canonical model using arguments typical of modal filtrations. All these results become relevant in a specific line of research where proof systems of deontic logic are intended to help in the testing of ethical behavior of AI through theorem-proving and model-checking (see \cite{arkoudas2005toward}, \cite{MUR}, \cite{bringsjord2006toward}).

Unfortunately, the Appendix is too long to include it here. Therefore, the reader can only find it at

\noindent\url{https://www.researchgate.net/publication/351656805_Appendix}.

\section{Conclusion}

``The performance is sometimes masterful, extremely clever, but the control of the actions, their source, is deranged and depends on various morbid impressions,'' says the character Zossimov, in Dostoevsky's \emph{Crime and Punishment}. From the discussions that appear in this paper, it is clear that what agents know and what they believe at the moment of acting---as well as the obligations that arise according to these knowledge and beliefs---can be interpreted as ``sources'' of their actions, as some of those ``impressions'' on which agency depends that Zossimov speaks about. 

The main novelty of the present, logic-based, treatment of these ``sources'' of agency lies in the incorporation of beliefs into deontic stit theory. The relation between (a) a given agent's doxastic state, (b) the actions that are available to said agent at some point of time, and (c) the deontic utility of such actions, gives us the opportunity to reason about a sense of agentive obligation that is based on the idea of maximizing expected deontic utility. Thus, we end up with a reasonable measure for explaining why agents could have favored certain actions over others, something that is useful in formal analyses of responsibility attribution, for instance.

A prominent feature of our analysis is the use of conditional beliefs. We mentioned that, since stit theory's account of interactive scenarios would greatly benefit from adopting viewpoints typical of EGT and of epistemic logic, we wanted to introduce a notion of belief that would satisfactorily open up possibilities for belief revision. It must be said, then, that the example discussed in Section \ref{giventofly} does not make heavy use of the theory of belief revision underlying the probabilistic semantics of belief that we introduced.  Although this was a choice made more for the sake of simplicity than anything else, it is true that the logic presented here is rather a `first step' toward an appropriate theory of belief-based action and obligation---a theory that would admit revision in \emph{both} the categories of beliefs and obligations. A very interesting problem for future work along these lines, then, regards implementing the ideas of belief revision---in terms of conditional belief---to formalize conditional doxastic oughts. The basic intuition is that, if at some index an agent has learned that $\psi$ is the case, then the doxastic obligations that such an agent had at the index should in principle be subject to the revision with $\psi$---just as beliefs are. Formulas of the form $\odot_\mathcal{B}[\alpha]^\psi \phi$ could then capture these revised doxastic oughts, such that possible semantics for these formulas could depend on the restriction of the model's domain to indices where $\psi$ holds---just as happens for the version of conditional belief discussed in this paper. In fact, one can find good pointers in this respect in \cite[Chapter 4]{Horty2001}, since a stit-theoretic account of conditional ought-to-do's is presented there. An adequate axiomatization of such possible conditional doxastic oughts, however, is still a complicated open problem. 

In conclusion, this work deals with important questions in the modeling of agency, knowledge, belief, and obligation. We presented logic-based characterizations of these concepts, that allowed us to devise unequivocal representations of interactive scenarios, where agents within an environment choose courses of action through time and where those choices could be traced back both to the epistemic-doxastic states of the agents and to different senses of obligation. The logic developed here lays the groundwork for interesting future research, and there is still plenty of work to do.






\bibliographystyle{eptcs} 
\bibliography{sample}


\end{document}

%% file: PaperPicture.tex
\usepackage{tikz}
\usetikzlibrary{calc,matrix,positioning,fit, shapes, automata, arrows, patterns}

\tikzset{
  solid node/.style={circle,draw,inner sep=1.2,fill=black},
  empty node/.style={font=\tiny, circle, outer sep=0pt, inner sep=0pt},
  matrix node/.style={rectangle, inner sep=0pt, draw, minimum height=0.5cm, minimum width=0.8cm, outer sep=0pt},  
  noline/.style={edge from parent/.style={draw=none}},
line/.style={edge from parent/.style={draw}},
}

%% file: main.bbl
\begin{thebibliography}{10}
\providecommand{\bibitemdeclare}[2]{}
\providecommand{\surnamestart}{}
\providecommand{\surnameend}{}
\providecommand{\urlprefix}{Available at }
\providecommand{\url}[1]{\texttt{#1}}
\providecommand{\href}[2]{\texttt{#2}}
\providecommand{\urlalt}[2]{\href{#1}{#2}}
\providecommand{\doi}[1]{doi:\urlalt{http://dx.doi.org/#1}{#1}}
\providecommand{\bibinfo}[2]{#2}

\bibitemdeclare{inproceedings}{abarca2019logic}
\bibitem{abarca2019logic}
\bibinfo{author}{Aldo Iv{\'a}n~Ram{\'\i}rez \surnamestart Abarca\surnameend} \&
  \bibinfo{author}{Jan \surnamestart Broersen\surnameend}
  (\bibinfo{year}{2019}): \emph{\bibinfo{title}{A Logic of Objective and
  Subjective Oughts}}.
\newblock In: {\sl \bibinfo{booktitle}{European Conference on Logics in
  Artificial Intelligence}}, \bibinfo{organization}{Springer}, pp.
  \bibinfo{pages}{629--641}, \doi{10.1007/978-3-030-19570-0\_41}.

\bibitemdeclare{inproceedings}{abarca2019stit}
\bibitem{abarca2019stit}
\bibinfo{author}{Aldo Iv{\'a}n~Ram{\'\i}rez \surnamestart Abarca\surnameend} \&
  \bibinfo{author}{Jan \surnamestart Broersen\surnameend}
  (\bibinfo{year}{2019}): \emph{\bibinfo{title}{Stit Semantics for Epistemic
  Notions Based on Information Disclosure in Interactive Settings}}.
\newblock In: {\sl \bibinfo{booktitle}{International Workshop on Dynamic
  Logic}}, \bibinfo{organization}{Springer}, pp. \bibinfo{pages}{171--189},
  \doi{10.1007/978-3-030-38808-9\_11}.

\bibitemdeclare{inproceedings}{arkoudas2005toward}
\bibitem{arkoudas2005toward}
\bibinfo{author}{Konstantine \surnamestart Arkoudas\surnameend},
  \bibinfo{author}{Selmer \surnamestart Bringsjord\surnameend} \&
  \bibinfo{author}{Paul \surnamestart Bello\surnameend} (\bibinfo{year}{2005}):
  \emph{\bibinfo{title}{Toward ethical robots via mechanized deontic logic}}.
\newblock In: {\sl \bibinfo{booktitle}{AAAI Fall Symposium on Machine Ethics}},
  pp. \bibinfo{pages}{17--23}.

\bibitemdeclare{article}{baltag2006conditional}
\bibitem{baltag2006conditional}
\bibinfo{author}{Alexandru \surnamestart Baltag\surnameend} \&
  \bibinfo{author}{Sonja \surnamestart Smets\surnameend}
  (\bibinfo{year}{2006}): \emph{\bibinfo{title}{Conditional doxastic models: A
  qualitative approach to dynamic belief revision}}.
\newblock {\sl \bibinfo{journal}{Electronic notes in theoretical computer
  science}} \bibinfo{volume}{165}, pp. \bibinfo{pages}{5--21},
  \doi{10.1016/j.entcs.2006.05.034}.

\bibitemdeclare{inproceedings}{baltag2006logic}
\bibitem{baltag2006logic}
\bibinfo{author}{Alexandru \surnamestart Baltag\surnameend} \&
  \bibinfo{author}{Sonja \surnamestart Smets\surnameend}
  (\bibinfo{year}{2006}): \emph{\bibinfo{title}{The logic of conditional
  doxastic actions: a theory of dynamic multi-agent belief revision}}.
\newblock In: {\sl \bibinfo{booktitle}{Proceedings of ESSLLI Workshop on
  Rationality and Knowledge}}, pp. \bibinfo{pages}{13--30}.

\bibitemdeclare{article}{baltag2008probabilistic}
\bibitem{baltag2008probabilistic}
\bibinfo{author}{Alexandru \surnamestart Baltag\surnameend} \&
  \bibinfo{author}{Sonja \surnamestart Smets\surnameend}
  (\bibinfo{year}{2008}): \emph{\bibinfo{title}{Probabilistic dynamic belief
  revision}}.
\newblock {\sl \bibinfo{journal}{Synthese}}
  \bibinfo{volume}{165}(\bibinfo{number}{2}), p. \bibinfo{pages}{179},
  \doi{10.1007/s11229-008-9369-8}.

\bibitemdeclare{book}{belnap01facing}
\bibitem{belnap01facing}
\bibinfo{author}{N.~\surnamestart Belnap\surnameend},
  \bibinfo{author}{M.~\surnamestart Perloff\surnameend} \&
  \bibinfo{author}{M.~\surnamestart Xu\surnameend} (\bibinfo{year}{2001}):
  \emph{\bibinfo{title}{Facing the future: agents and choices in our
  indeterminist world}}.
\newblock \bibinfo{publisher}{Oxford University Press}.

\bibitemdeclare{article}{bjorndahl2017reasoning}
\bibitem{bjorndahl2017reasoning}
\bibinfo{author}{Adam \surnamestart Bjorndahl\surnameend},
  \bibinfo{author}{Joseph~Y \surnamestart Halpern\surnameend} \&
  \bibinfo{author}{Rafael \surnamestart Pass\surnameend}
  (\bibinfo{year}{2017}): \emph{\bibinfo{title}{Reasoning about rationality}}.
\newblock {\sl \bibinfo{journal}{Games and Economic Behavior}}
  \bibinfo{volume}{104}, pp. \bibinfo{pages}{146--164},
  \doi{10.1016/j.geb.2017.03.006}.

\bibitemdeclare{article}{board2004dynamic}
\bibitem{board2004dynamic}
\bibinfo{author}{Oliver \surnamestart Board\surnameend} (\bibinfo{year}{2004}):
  \emph{\bibinfo{title}{Dynamic interactive epistemology}}.
\newblock {\sl \bibinfo{journal}{Games and Economic Behavior}}
  \bibinfo{volume}{49}(\bibinfo{number}{1}), pp. \bibinfo{pages}{49--80},
  \doi{10.1016/j.geb.2003.10.006}.

\bibitemdeclare{article}{boutilier1995revision}
\bibitem{boutilier1995revision}
\bibinfo{author}{Craig \surnamestart Boutilier\surnameend} et~al.
  (\bibinfo{year}{1995}): \emph{\bibinfo{title}{On the revision of
  probabilistic belief states}}.
\newblock {\sl \bibinfo{journal}{Notre Dame Journal of Formal Logic}}
  \bibinfo{volume}{36}(\bibinfo{number}{1}), pp. \bibinfo{pages}{158--183},
  \doi{10.1305/ndjfl/1040308833}.

\bibitemdeclare{article}{bringsjord2006toward}
\bibitem{bringsjord2006toward}
\bibinfo{author}{Selmer \surnamestart Bringsjord\surnameend},
  \bibinfo{author}{Konstantine \surnamestart Arkoudas\surnameend} \&
  \bibinfo{author}{Paul \surnamestart Bello\surnameend} (\bibinfo{year}{2006}):
  \emph{\bibinfo{title}{Toward a general logicist methodology for engineering
  ethically correct robots}}.
\newblock {\sl \bibinfo{journal}{IEEE Intelligent Systems}}
  \bibinfo{volume}{21}(\bibinfo{number}{4}), pp. \bibinfo{pages}{38--44},
  \doi{10.1109/MIS.2006.82}.

\bibitemdeclare{article}{broersen2013probabilistic}
\bibitem{broersen2013probabilistic}
\bibinfo{author}{Jan \surnamestart Broersen\surnameend} (\bibinfo{year}{2013}):
  \emph{\bibinfo{title}{Probabilistic stit logic and its decomposition}}.
\newblock {\sl \bibinfo{journal}{International journal of approximate
  reasoning}} \bibinfo{volume}{54}(\bibinfo{number}{4}), pp.
  \bibinfo{pages}{467--477}, \doi{10.1016/j.ijar.2012.08.007}.

\bibitemdeclare{inproceedings}{JANANDI}
\bibitem{JANANDI}
\bibinfo{author}{Jan \surnamestart Broersen\surnameend} \&
  \bibinfo{author}{Aldo~Iv{\'a}n \surnamestart Ram{\'\i}rez~Abarca\surnameend}
  (\bibinfo{year}{2018}): \emph{\bibinfo{title}{Formalising Oughts and
  Practical Knowledge without Resorting to Action Types}}.
\newblock In: {\sl \bibinfo{booktitle}{Proceedings of the 17th International
  Conference on Autonomous Agents and MultiAgent Systems}},
  \bibinfo{organization}{International Foundation for Autonomous Agents and
  Multiagent Systems}, pp. \bibinfo{pages}{1877--1879}.

\bibitemdeclare{book}{de1936probabilites}
\bibitem{de1936probabilites}
\bibinfo{author}{Bruno \surnamestart De~Finetti\surnameend}
  (\bibinfo{year}{1936}): \emph{\bibinfo{title}{Les probabilit{\'e}s nulles}}.
\newblock \bibinfo{publisher}{Gauthier-Villars}.

\bibitemdeclare{article}{duijf2016representing}
\bibitem{duijf2016representing}
\bibinfo{author}{Hein \surnamestart Duijf\surnameend} \& \bibinfo{author}{Jan
  \surnamestart Broersen\surnameend} (\bibinfo{year}{2016}):
  \emph{\bibinfo{title}{Representing strategies}}.
\newblock {\sl \bibinfo{journal}{arXiv preprint arXiv:1607.03355}},
  \doi{10.4204/EPTCS.218.2}.

\bibitemdeclare{phdthesis}{duijf2018let}
\bibitem{duijf2018let}
\bibinfo{author}{HWA \surnamestart Duijf\surnameend} (\bibinfo{year}{2018}):
  \emph{\bibinfo{title}{Let's do it!: Collective responsibility, joint action,
  and participation}}.
\newblock Ph.D. thesis, \bibinfo{school}{Utrecht University}.

\bibitemdeclare{phdthesis}{gkikas2015stable}
\bibitem{gkikas2015stable}
\bibinfo{author}{Konstantinos \surnamestart Gkikas\surnameend}
  (\bibinfo{year}{2015}): \emph{\bibinfo{title}{Stable Beliefs and Conditional
  Probability Spaces}}.
\newblock Ph.D. thesis, \bibinfo{school}{Universiteit van Amsterdam}.

\bibitemdeclare{article}{halpern2010lexicographic}
\bibitem{halpern2010lexicographic}
\bibinfo{author}{Joseph~Y \surnamestart Halpern\surnameend}
  (\bibinfo{year}{2010}): \emph{\bibinfo{title}{Lexicographic probability,
  conditional probability, and nonstandard probability}}.
\newblock {\sl \bibinfo{journal}{Games and Economic Behavior}}
  \bibinfo{volume}{68}(\bibinfo{number}{1}), pp. \bibinfo{pages}{155--179},
  \doi{10.1016/j.geb.2009.03.013}.

\bibitemdeclare{incollection}{hammond1994elementary}
\bibitem{hammond1994elementary}
\bibinfo{author}{Peter~J \surnamestart Hammond\surnameend}
  (\bibinfo{year}{1994}): \emph{\bibinfo{title}{Elementary non-Archimedean
  representations of probability for decision theory and games}}.
\newblock In: {\sl \bibinfo{booktitle}{Patrick Suppes: scientific
  philosopher}}, \bibinfo{publisher}{Springer}, pp. \bibinfo{pages}{25--61},
  \doi{10.1007/978-94-011-0774-7\_2}.

\bibitemdeclare{article}{harsanyi1967games}
\bibitem{harsanyi1967games}
\bibinfo{author}{John~C \surnamestart Harsanyi\surnameend}
  (\bibinfo{year}{1967}): \emph{\bibinfo{title}{Games with incomplete
  information played by “Bayesian” players, I--III Part I. The basic
  model}}.
\newblock {\sl \bibinfo{journal}{Management science}}
  \bibinfo{volume}{14}(\bibinfo{number}{3}), pp. \bibinfo{pages}{159--182},
  \doi{10.1287/mnsc.14.3.159}.

\bibitemdeclare{article}{horty2019epistemic}
\bibitem{horty2019epistemic}
\bibinfo{author}{John \surnamestart Horty\surnameend} (\bibinfo{year}{2019}):
  \emph{\bibinfo{title}{Epistemic Oughts in Stit Semantics}}.
\newblock {\sl \bibinfo{journal}{Ergo, an Open Access Journal of Philosophy}}
  \bibinfo{volume}{6}, \doi{10.3998/ergo.12405314.0006.004}.

\bibitemdeclare{article}{horty2017action}
\bibitem{horty2017action}
\bibinfo{author}{John \surnamestart Horty\surnameend} \& \bibinfo{author}{Eric
  \surnamestart Pacuit\surnameend} (\bibinfo{year}{2017}):
  \emph{\bibinfo{title}{Action types in stit semantics}}.
\newblock {\sl \bibinfo{journal}{The Review of Symbolic Logic}}
  \bibinfo{volume}{10}(\bibinfo{number}{4}), pp. \bibinfo{pages}{617--637},
  \doi{10.1017/S1755020317000016}.

\bibitemdeclare{book}{Horty2001}
\bibitem{Horty2001}
\bibinfo{author}{John~F. \surnamestart Horty\surnameend}
  (\bibinfo{year}{2001}): \emph{\bibinfo{title}{Agency and Deontic Logic}}.
\newblock \bibinfo{publisher}{Oxford University Press},
  \doi{10.1093/0195134613.001.0001}.

\bibitemdeclare{article}{jeffrey1965ethics}
\bibitem{jeffrey1965ethics}
\bibinfo{author}{Richard~C \surnamestart Jeffrey\surnameend}
  (\bibinfo{year}{1965}): \emph{\bibinfo{title}{Ethics and the Logic of
  Decision}}.
\newblock {\sl \bibinfo{journal}{The Journal of Philosophy}}
  \bibinfo{volume}{62}(\bibinfo{number}{19}), pp. \bibinfo{pages}{528--539},
  \doi{10.2307/2023748}.

\bibitemdeclare{incollection}{karni2014axiomatic}
\bibitem{karni2014axiomatic}
\bibinfo{author}{Edi \surnamestart Karni\surnameend} (\bibinfo{year}{2014}):
  \emph{\bibinfo{title}{Axiomatic foundations of expected utility and
  subjective probability}}.
\newblock In: {\sl \bibinfo{booktitle}{Handbook of the Economics of Risk and
  Uncertainty}}, \bibinfo{volume}{1}, \bibinfo{publisher}{Elsevier}, pp.
  \bibinfo{pages}{1--39}, \doi{10.1016/B978-0-444-53685-3.00001-5}.

\bibitemdeclare{article}{lehmann1992does}
\bibitem{lehmann1992does}
\bibinfo{author}{Daniel \surnamestart Lehmann\surnameend} \&
  \bibinfo{author}{Menachem \surnamestart Magidor\surnameend}
  (\bibinfo{year}{1992}): \emph{\bibinfo{title}{What does a conditional
  knowledge base entail?}}
\newblock {\sl \bibinfo{journal}{Artificial intelligence}}
  \bibinfo{volume}{55}(\bibinfo{number}{1}), pp. \bibinfo{pages}{1--60},
  \doi{10.1016/0004-3702(92)90041-U}.

\bibitemdeclare{article}{lorini2014logical}
\bibitem{lorini2014logical}
\bibinfo{author}{Emiliano \surnamestart Lorini\surnameend},
  \bibinfo{author}{Dominique \surnamestart Longin\surnameend} \&
  \bibinfo{author}{Eunate \surnamestart Mayor\surnameend}
  (\bibinfo{year}{2014}): \emph{\bibinfo{title}{A logical analysis of
  responsibility attribution: emotions, individuals and collectives}}.
\newblock {\sl \bibinfo{journal}{Journal of Logic and Computation}}
  \bibinfo{volume}{24}(\bibinfo{number}{6}), pp. \bibinfo{pages}{1313--1339},
  \doi{10.1093/logcom/ext072}.

\bibitemdeclare{article}{MUR}
\bibitem{MUR}
\bibinfo{author}{Yuko \surnamestart Murakami\surnameend}
  (\bibinfo{year}{2004}): \emph{\bibinfo{title}{Utilitarian deontic logic}}.
\newblock {\sl \bibinfo{journal}{AiML-2004: Advances in Modal Logic}}
  \bibinfo{volume}{287}.

\bibitemdeclare{incollection}{sep-epistemic-game}
\bibitem{sep-epistemic-game}
\bibinfo{author}{Eric \surnamestart Pacuit\surnameend} \&
  \bibinfo{author}{Olivier \surnamestart Roy\surnameend}
  (\bibinfo{year}{2017}): \emph{\bibinfo{title}{{Epistemic Foundations of Game
  Theory}}}.
\newblock In \bibinfo{editor}{Edward~N. \surnamestart Zalta\surnameend},
  editor: {\sl \bibinfo{booktitle}{The {Stanford} Encyclopedia of Philosophy}},
  \bibinfo{edition}{{S}ummer 2017} edition, \bibinfo{publisher}{Metaphysics
  Research Lab, Stanford University}.

\bibitemdeclare{book}{peterson2017introduction}
\bibitem{peterson2017introduction}
\bibinfo{author}{Martin \surnamestart Peterson\surnameend}
  (\bibinfo{year}{2017}): \emph{\bibinfo{title}{An introduction to decision
  theory}}.
\newblock \bibinfo{publisher}{Cambridge University Press},
  \doi{10.1017/9781316585061}.

\bibitemdeclare{book}{popper1968logic}
\bibitem{popper1968logic}
\bibinfo{author}{Karl~Raimund \surnamestart Popper\surnameend}
  (\bibinfo{year}{1968}): \emph{\bibinfo{title}{The Logic of Scientific
  Discovery.(Revised Edition.).}}
\newblock \bibinfo{publisher}{Hutchinson}.

\bibitemdeclare{article}{renyi1955new}
\bibitem{renyi1955new}
\bibinfo{author}{Alfr{\'e}d \surnamestart R{\'e}nyi\surnameend}
  (\bibinfo{year}{1955}): \emph{\bibinfo{title}{On a new axiomatic theory of
  probability}}.
\newblock {\sl \bibinfo{journal}{Acta Mathematica Academiae Scientiarum
  Hungarica}} \bibinfo{volume}{6}(\bibinfo{number}{3-4}), pp.
  \bibinfo{pages}{285--335}, \doi{10.1007/BF02024393}.

\bibitemdeclare{article}{robinson1973function}
\bibitem{robinson1973function}
\bibinfo{author}{Abraham \surnamestart Robinson\surnameend}
  (\bibinfo{year}{1973}): \emph{\bibinfo{title}{Function theory on some
  nonarchimedean fields}}.
\newblock {\sl \bibinfo{journal}{The American Mathematical Monthly}}
  \bibinfo{volume}{80}(\bibinfo{number}{6}), pp. \bibinfo{pages}{87--109},
  \doi{10.2307/3038223}.

\bibitemdeclare{book}{Savage1954}
\bibitem{Savage1954}
\bibinfo{author}{L.J. \surnamestart Savage\surnameend} (\bibinfo{year}{1954}):
  \emph{\bibinfo{title}{The Foundations of Statistics}}.
\newblock \bibinfo{publisher}{John Wiley and Sons}, \bibinfo{address}{New
  York}.

\bibitemdeclare{article}{sharpe2018risk}
\bibitem{sharpe2018risk}
\bibinfo{author}{Keiran \surnamestart Sharpe\surnameend}
  (\bibinfo{year}{2018}): \emph{\bibinfo{title}{On risk and uncertainty, and
  objective versus subjective probability}}.
\newblock {\sl \bibinfo{journal}{Economic Record}} \bibinfo{volume}{94}, pp.
  \bibinfo{pages}{49--72}, \doi{10.1111/1475-4932.12403}.

\bibitemdeclare{incollection}{sep-decision-theory}
\bibitem{sep-decision-theory}
\bibinfo{author}{Katie \surnamestart Steele\surnameend} \&
  \bibinfo{author}{H.~Orri \surnamestart Stefánsson\surnameend}
  (\bibinfo{year}{2016}): \emph{\bibinfo{title}{Decision Theory}}.
\newblock In \bibinfo{editor}{Edward~N. \surnamestart Zalta\surnameend},
  editor: {\sl \bibinfo{booktitle}{The Stanford Encyclopedia of Philosophy}},
  \bibinfo{edition}{winter 2016} edition, \bibinfo{publisher}{Metaphysics
  Research Lab, Stanford University}.

\bibitemdeclare{article}{tamminga2013deontic}
\bibitem{tamminga2013deontic}
\bibinfo{author}{Allard \surnamestart Tamminga\surnameend}
  (\bibinfo{year}{2013}): \emph{\bibinfo{title}{Deontic logic for strategic
  games}}.
\newblock {\sl \bibinfo{journal}{Erkenntnis}}
  \bibinfo{volume}{78}(\bibinfo{number}{1}), pp. \bibinfo{pages}{183--200},
  \doi{10.1007/s10670-011-9349-0}.

\bibitemdeclare{article}{van1995fine}
\bibitem{van1995fine}
\bibinfo{author}{Bas~C \surnamestart Van~Fraassen\surnameend}
  (\bibinfo{year}{1995}): \emph{\bibinfo{title}{Fine-grained opinion,
  probability, and the logic of full belief}}.
\newblock {\sl \bibinfo{journal}{Journal of Philosophical logic}}
  \bibinfo{volume}{24}(\bibinfo{number}{4}), pp. \bibinfo{pages}{349--377},
  \doi{10.1007/BF01048352}.

\bibitemdeclare{article}{wansing2006doxastic}
\bibitem{wansing2006doxastic}
\bibinfo{author}{Heinrich \surnamestart Wansing\surnameend}
  (\bibinfo{year}{2006}): \emph{\bibinfo{title}{Doxastic decisions, epistemic
  justification, and the logic of agency}}.
\newblock {\sl \bibinfo{journal}{Philosophical Studies}}
  \bibinfo{volume}{128}(\bibinfo{number}{1}), pp. \bibinfo{pages}{201--227},
  \doi{10.1007/s11098-005-4063-x}.

\bibitemdeclare{inproceedings}{xu1994decidability}
\bibitem{xu1994decidability}
\bibinfo{author}{Ming \surnamestart Xu\surnameend} (\bibinfo{year}{1994}):
  \emph{\bibinfo{title}{Decidability of deliberative stit theories with
  multiple agents}}.
\newblock In: {\sl \bibinfo{booktitle}{International Conference on Temporal
  Logic}}, \bibinfo{organization}{Springer}, pp. \bibinfo{pages}{332--348},
  \doi{10.1007/BFb0013997}.

\bibitemdeclare{article}{Xu2015}
\bibitem{Xu2015}
\bibinfo{author}{Ming \surnamestart Xu\surnameend} (\bibinfo{year}{2015}):
  \emph{\bibinfo{title}{Combinations of Stit with Ought and Know}}.
\newblock {\sl \bibinfo{journal}{Journal of Philosophical Logic}}
  \bibinfo{volume}{44}(\bibinfo{number}{6}), pp. \bibinfo{pages}{851--877},
  \doi{10.1007/s10992-015-9365-7}.

\end{thebibliography}
